\documentclass{aa}  

\usepackage{ulem}
\usepackage{color}

\usepackage{graphicx}
\usepackage{txfonts}
\usepackage{amsmath}
\newcommand{\vv}[1]{\mathbf{#1}}

\begin{document}

   \title{Solar wind dynamics around a comet}
   \subtitle{A 2D semi-analytical kinetic model}

   \author{E. Behar
          \inst{1,2},
          B. Tabone
          \inst{3},
          M. Saillenfest
          \inst{4},
          P. Henri
          \inst{5},
          J. Deca
          \inst{6,7},
          J. Lindkvist
          \inst{8},
          M. Holmstr\"om
          \inst{1},
          H. Nilsson
          \inst{1}}
   \authorrunning{E. Behar et al.}
          
   \institute{Swedish Institute of Space Physics, Kiruna, Sweden.\and
               Lule\aa\ University of Technology, Department of Computer Science, Electrical and Space Engineering, Kiruna, Sweden\and
               LERMA, Observatoire de Paris, PSL Research University, CNRS, Sorbonne Universit\'e, UPMC Univ. Paris 06, 75014 Paris, France \and
               IMCCE, Observatoire de Paris, PSL Research University, CNRS, Sorbonne Universit\'e, UPMC Univ. Paris 06, LAL, Universit{\'e} de Lille, 75014 Paris, France\and
               LPC2E, CNRS, Orl\'eans, France\and
               Laboratory for Atmospheric and Space Physics (LASP), University of Colorado Boulder, Boulder, Colorado 80303, USA\and 
               Institute for Modeling Plasma, Atmospheres and Cosmic Dust (IMPACT), NASA/SSERVI, Boulder, Colorado 80303, USA\and 
               Ume\aa University, Department of Physics, Ume\aa, Sweden \\
              \email{etienne.behar@irf.se}             }

   \date{Received 2018-01-31; accepted 2018-04-28}
 
  \abstract
   {}
   {We aim at analytically modelling the solar wind proton trajectories during their interaction with a partially ionised cometary atmosphere, not in terms of bulk properties of the flow but in terms of single particle dynamics.}
   {We first derive a generalised gyromotion, in which the electric field is reduced to its motional component. Steady-state is assumed, and simplified models of the cometary density and of the electron fluid are used to express the force experienced by individual solar wind protons during the interaction. }
   {A three-dimensional (3D) analytical expression of the gyration of two interacting plasma beams is obtained. Applying it to a comet case, the force on protons is always perpendicular to their velocity and has an amplitude proportional to $1/r^2$~. The solar wind deflection is obtained at any point in space. The resulting picture presents a caustic of intersecting trajectories, and a circular region is found that is completely free of particles. The particles do not lose any kinetic energy and this absence of deceleration, together with the solar wind deflection pattern and the presence of a solar wind ion cavity, is in good agreement with the general results of the {\it Rosetta} mission.}
   {The qualitative match between the model and the {\it in situ} data highlights how dominant the motional electric field is throughout most of the interaction region for the solar wind proton dynamics. The model provides a simple general kinetic description of how momentum is transferred between these two collisionless plasmas. It also shows the potential of this semi-analytical model for a systematic quantitative comparison to the data.}

   \keywords{Comets: general, Methods: analytical, Plasmas}

\maketitle

\section{A global model}

        The plasma interaction between the solar wind and a cometary atmosphere (coma) offers a unique situation in the solar system. The absence of an intrinsic magnetic field, the typical small size of the nucleus, and its negligible gravity combined with its highly elliptical orbit result in an ever changing interaction, in which the coma continuously and completely escapes the comet, dragged away by the magnetised stream of solar particles. These properties also result in one of the largest obstacles to the solar wind in the solar system. At comet Halley, the first cometary ions were detected 7.8 million kilometres away from the nucleus by the {\it Giotto} probe \citep{johnstone1986nature}, a distance comparable to the day-side extent of the Jovian magnetosphere.
        
        A major advance to comprehend this interaction was proposed by \citet{alfven1957tellus}, who emphasised the role of the solar wind magnetic field in the formation of the cometary tails. \citet{biermann1967sp} proposed a model of the day-side of a comet atmosphere in a hydrodynamical description of the interaction. These and all the previous efforts were tackling the features of the comet's head and tails that were visible from Earth, naturally directing the scientific interest towards strongly outgassing comets close to their perihelion. In the next two decades, space probes were leaving Earth targeting such active comets, and at Giacobini-Zinner and Halley, what had previously been invisible became visible: a whole set of plasma structures came within the reach of scientists \citep{halley1988springer, cowley1987ptrsl, gombosi2015}.
        
        Between 2014 and 2016, the {\it Rosetta} spacecraft cohabited for more than two years with its host body, comet 67P/Churyumov-Gerasimenko (67P/CG), enabling for the first time observations at large heliocentric distances (> 3 au). Scientists were given the opportunity to witness the early interaction between a young tenuous coma and the solar wind, far away from the Sun. The nature of such an interaction was entirely new. Indeed, whereas at previously visited comets the interaction region was much larger than the scale of the ion gyromotion, resulting in what one could call a "fluid comet", for which the classical fluid treatment of the plasmas applies, at 67P/CG and at large heliocentric distances, the ion transit timescale through the coma is shorter than its gyroperiod, resulting in a "kinetic comet" for which no analytical approach is available so far. 
        
        Using \textit{in situ} measurements, the evolution of this interaction was followed carefully \citep{behar2016aa, nilsson2017mnras} resulting in some surprising findings. Initially barely disturbed, the solar wind started displaying a peculiar behaviour as the nucleus was getting closer to the Sun. Its flow slowly diverged from the Sun-comet direction, to eventually be seen flowing almost back towards the Sun at speeds of hundreds of kilometres per second. Eventually, the flow vanished from the {\it in situ} measurements: a void of solar ions was formed around the nucleus, while no severe deceleration was observed \citep{behar2017mnras}. The same interaction -- for large to intermediate heliocentric distances -- was also tackled by several simulation efforts, using hybrid particle-in-cell models \citep{bagdonat2002emp, hansen2007ssr, koenders2016aa, koenders2016mnras, behar2016aa}, as well as a fully kinetic model \citep{deca2017prl}. At heliocentric distances down to less than 2 au, simulations result in a highly asymmetric plasma environment. In particular, the solar wind presents a structure of high ion density only seen in the hemisphere of the coma opposite to the direction of the upstream electric field, one of the typical signatures of such a kinetic comet. This structure is interpreted as a Mach cone by \citet{bagdonat2002emp}, a term adopted in several of the cited simulation studies. However until now, this asymmetric structure was only found in numerical models with intricate physics, and generally lacks the physical interpretation which would elucidate the experimental results exposed above.
        
        Our goal is to understand and analyse the mechanism which transfers momentum between the solar wind and the coma, leading to such a deflection of a barely decelerated flow, in the absence of collisions. Additionally, we aim at providing novel insight into the nature of this asymmetric solar wind structure, which may be considered as the very seed of a cometary magnetosphere. \\

        The present attempt to model the interaction puts the emphasis on the role of the motional electric field by considering the parameter space region, in which currents orthogonal to the magnetic field and pressure gradients can be neglected. Under these conditions, any noticeable disturbances of the flow necessarily result from the integrated interaction with and through the smooth and extended obstacle. This is in contrast to the situation at more classical and massive solar system bodies with intrinsic magnetic fields, dense and limited atmospheres, or conductive cores inducing a magnetic feedback to the solar wind. There, contrasted plasma structures and boundaries are formed, such as bow shocks, magnetopauses, induced magnetospheric boundaries or ionopauses. The obstacle to the solar flow is therefore compact and localised. Similar plasma boundaries may also appear at comets. For example, weak bow shocks were observed at comets Halley, Giacobini-Zinner, Grigg-Skjellerup and Borrelly \citep{coates2009aip}. However, such boundaries are only formed close to the Sun, and even in that case, the neutral atmosphere extends further out (a weak bow shock was observed about a million kilometres away from comet Halley's nucleus, after the detection of the first cometary ions \citep{johnstone1986nature}), and mass-loading (the addition of new-born cometary ions to the solar wind) takes place, whether boundaries are formed closer to the nucleus or not. Therefore the present model should be representative and relevant for the region beyond the potential bow shock which forms when a comet gets closer to the Sun.\\

        To infer the global behaviour of a system, whenever possible, an analytical model may overcome intrinsic limitations of {\it in situ} data (one-point measurement, instrumental errors and limitations) and simulation data (simulation of only a finite region of space, intricate physics, numerical limitations). While doing so, it allows to encapsulate one or a few of the driving mechanisms of a system in a reduced form, though at the cost of realism. In the present series of articles, the synergy between these three approaches -- experimental data, numerical simulations, and theoretical models --  is explored. This article focuses on the physical model and provides an expression for the force experienced by single solar wind protons, through the extended coma. The corresponding dynamics is thoroughly solved by \citet{saillenfest2018aa}, a solution widely used in the present work. The semi-analytical model we propose is computationally very cheap, and allows for a systematic comparison to each and every {\it in situ} data point. This extended comparison, together with the comparisons to numerical simulations, follows in subsequent articles of the series. \\
        
        The model developed in the following sections requires several sub-models. One is a description of the ionised coma and its density distribution, following the same need for simplicity-to-relevancy ratio. The second is a description of the electric field and the magnetic field, which piles up due to the local decrease in the average velocity of the electrons, as slow new-born cometary ions are added to the flow. The motional electric field is completely dependent on the motion of the particles, which itself depends on the electric and magnetic fields. This inter-dependency is tackled in the following section as a generalised gyromotion, and results in a three-dimensional (3D) model of its own.


\section{Generalised gyromotion}

        In this section, we derive the general dynamics of the interaction between two collisionless beams of plasma that are only subject to the Lorentz force. The subscripts $sw$ and $com$ are used for parameters and values of the solar wind and the cometary ion populations, respectively. The characteristic length, time, and velocity of the system are $\ell$~, $t$ and $u$~. $\vv{E}$ and $\vv{B}$ are the electric and the magnetic field, $\underline{{\vv u}_i}$ is the average velocity of all charges carried by ions, $n$ is the plasma number density, $e$ is the elementary charge, $\vv{j}$ is the electric current and $\vv{P}_e$ is the electron pressure.
        
        Our starting point is the simplified Ohm's law, in which the electron inertial and the resistivity/collisional terms are neglected (see, e.g. \citet{valentini2007jpc}). The plasma is weakly magnetised and quasi-neutral. The system is considered to be at steady state: $\partial_t \cdot \equiv 0$~.
        
        \begin{equation}
            {\vv E} = -\underline{{\vv u}_i} \times {\vv B} + \frac{1}{ne} \ \vv{j} \times \vv{B} - \frac{1}{ne} \vv{\nabla} \vv{P}_e
            \label{Egen}
        .\end{equation}
        
        The total electric field exhibits three distinct components: the motional electric field, the Hall term, and the pressure gradient term. Analysing orders of magnitude in these three terms, as well as in the Amp\`ere's law, the Faraday's law, and the Lorentz force, the following orderings can be found.
        
        \begin{equation}
        \begin{array}{rcl}
                \ell \gg d_i & \Rightarrow & |\underline{{\vv u}_i} \times {\vv B}| \gg 1/ne \; | \vv{j} \times \vv{B} | \\
                \ell_P \gg r_{ge} v_{the}/u & \Rightarrow & |\underline{{\vv u}_i} \times {\vv B}| \gg 1/ne \; | \vv{\nabla} \vv{P}_e |
                \label{scales}
        \end{array}
        .\end{equation}
        
        In these expressions, $d_i$ is the ion inertial length, and $\ell_P$ is the characteristic length of the pressure gradient. $r_{ge} = m_eu_e/(eB)$ is the electron gyroradius and $v_{the} = \sqrt{2kT_e/m_e}$~ the electron thermal speed. At these scales (Eq. (\ref{scales})), the electric field is reduced to
        
        \begin{equation}
            {\vv E} = -\underline{{\vv u}_i} \times {\vv B}
            \label{E}
        .\end{equation}

        This directly implies that the currents perpendicular to the magnetic field are negligible for scales $\ell \gg d_i$~. The average velocity $\underline{\vv{u}_i}$ in the case of our two beams can be reduced to only two terms:
        
        \begin{equation}
        \begin{array}{rcl}
            \underline{{\vv u}_i} & = & \xi_{sw}\underline{\vv{u}_{sw}} + \xi_{com}\underline{\vv{u}_{com}}\\
            &&\\
            \xi_{sw} & = & \dfrac{n_{sw}q_{sw} }{ n_{sw}q_{sw} + n_{com}q_{com}} ~;~ \xi_{com}  =  \dfrac{n_{com}q_{com} }{ n_{sw}q_{sw} + n_{com}q_{com}}
            \label{velIon}
        \end{array}
        .\end{equation}
        
        In the absence of any force  other than the Lorentz force, the dynamics of a single particle in either of the two populations is described by the following system of ordinary differential equations:
        
        \begin{equation}
        \begin{array}{|rcl}
             \dot{\vv u}_{sw} & = & q_{sw}/m_{sw} \left( {\vv E} + {\vv u}_{sw} \times {\vv B}\right) \\
             \dot{\vv u}_{com} & = & q_{com}/m_{com} \left( {\vv E} + {\vv u}_{com} \times {\vv B}\right)
        \end{array}
        \label{ODE}
        .\end{equation}
        
        Considering two initially perfect beams in velocity space, we have $\vv{u}_{com} = \underline{\vv{u}_{com}}$ and $\vv{u}_{sw} = \underline{\vv{u}_{sw}}$~: all particles of the population experience the same acceleration at the same time. The temperature is not defined.
        
        Without loss of generality, one can choose a frame in which the magnetic field is directed along the $y$-axis. With (\ref{E}), (\ref{velIon}), (\ref{ODE}),  and $q_{sw} = q_{com} = q$~:
        
        \begin{equation}
           \begin{array}{|rcl}
                \dot{\vv u}_{sw} & = & ~\dfrac{q \ \xi_{com} B}{m_{sw}} ( {\vv u}_{sw} - {\vv u}_{com} ) \times \hat{\vv y} \\
                ~ \\
                \dot{\vv u}_{com} & = &  -\dfrac{q \ \xi_{sw} B}{m_{com}} ( {\vv u}_{sw} - {\vv u}_{com} ) \times \hat{\vv y}
           \end{array}
           \label{dynamics}
        .\end{equation}
        
        A first important result is that there can be no particle acceleration along the magnetic field.
        
\begin{center}
        \begin{figure}
                \includegraphics[width=.9\columnwidth]{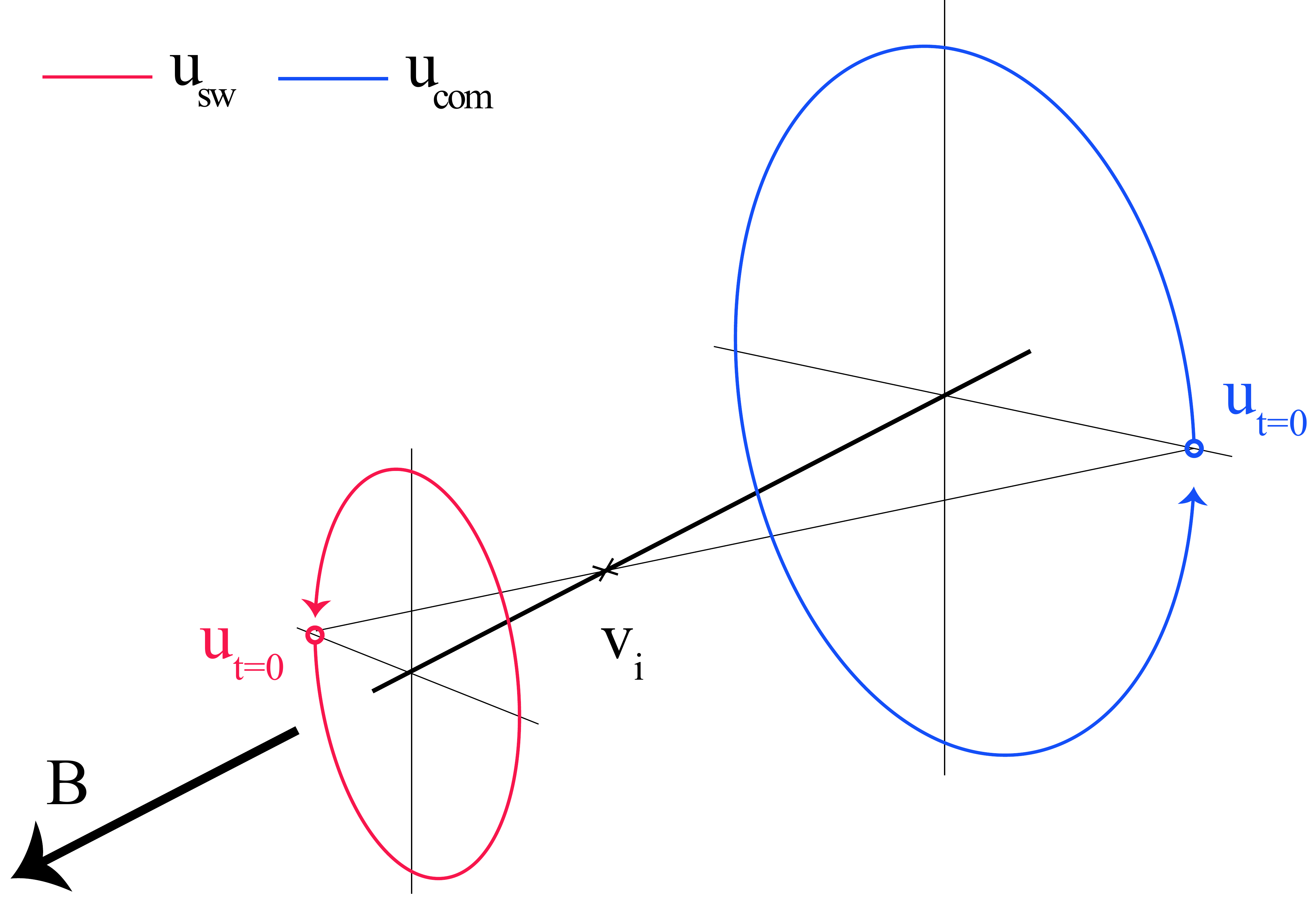}
                \caption{Evolution in velocity space of two interacting beams of plasma, for the most general configuration. \label{genGyr}}
        \end{figure}
        \end{center}

        A second noteworthy result is that the velocity of the centre of mass, defined as
        
        \begin{equation}
         \underline{\vv{v}_i} = \frac{n_{sw} m_{sw}}{n_{sw} m_{sw} + n_{com} m_{com}} \vv{u}_{sw} + \frac{n_{com} m_{com}}{n_{sw} m_{sw} + n_{com} m_{com}} \vv{u}_{com}
         \end{equation}
         
        is conserved through time, $\underline{\dot{\vv{v}}_i} = \vv{0}$~. This holds over spatial scales shorter than $\ell$ and $\ell_P$~, and if no mass is added. The equations of motion have the general form $\dot{\vv{u}} = \omega \ \vv{u} \times \hat{\vv{y}}$~. In velocity space, the two beams move along circles perpendicular to the magnetic field (no acceleration along the magnetic field), as shown in Fig. \ref{genGyr}.
        
        \begin{center}
        \begin{figure}
                \includegraphics[width=.9\columnwidth]{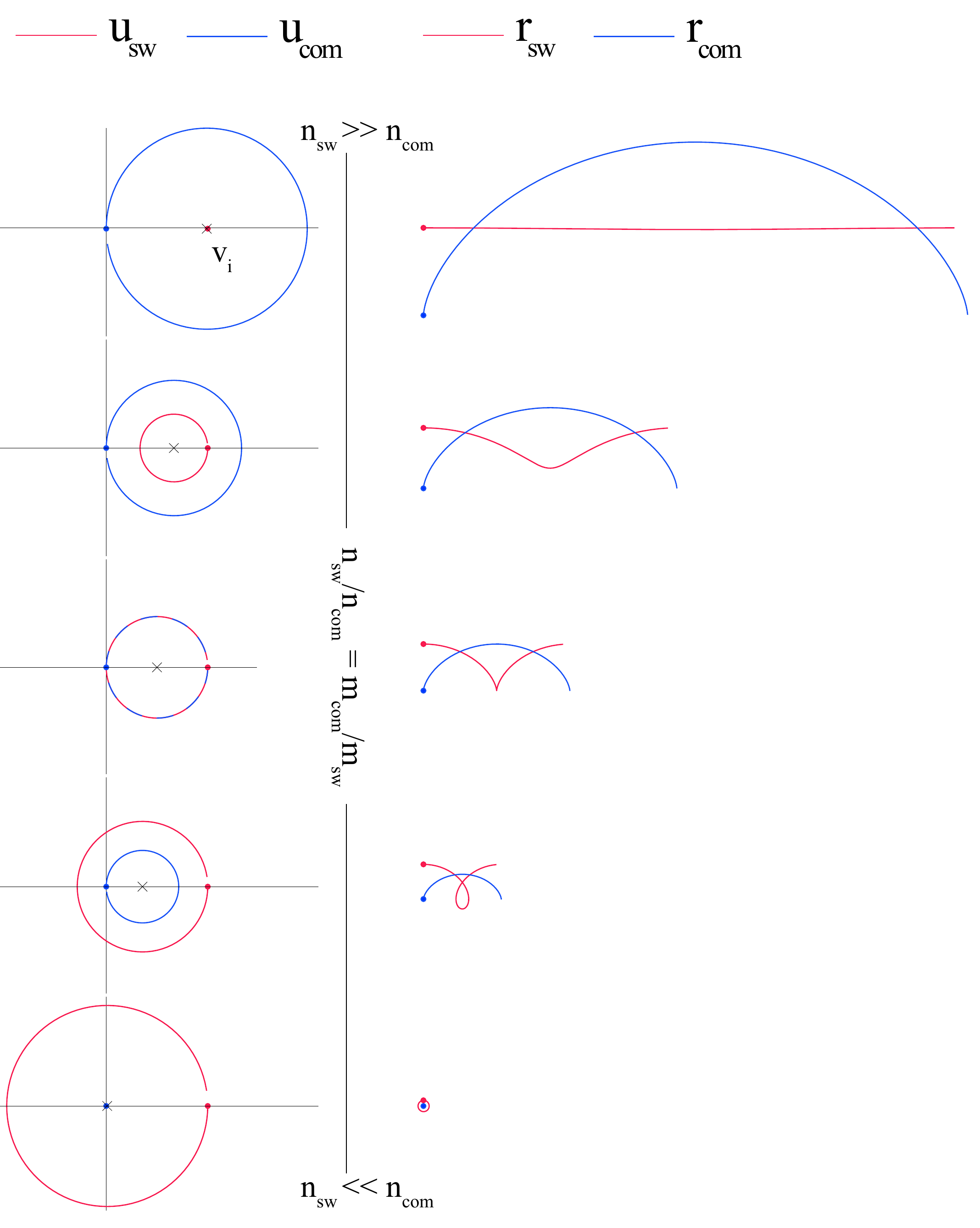}
                \caption{Evolution of both populations in velocity and physical space (left and right columns, respectively), projected in the plane perpendicular to the magnetic field in the comet frame for different density ratios (top to bottom). $\underline{\vv{v}_i}$ is shown with a black cross in the left column. \label{general}}
        \end{figure}
        \end{center}
        
         If the two beams have initially the same parallel velocity (velocity component along the magnetic field), the two circles are in one and the same plane, centred on $\underline{\vv{v}_i}$~, independent of any change of inertial frame. This can easily be seen in Fig. \ref{genGyr}. Then in the frame in which $\underline{\vv{v}_i}=0$~, the two populations describe circles in velocity {\it and} physical space. The generalised gyrofrequency and gyroradii are then:
        \begin{equation}
                \begin{array}{rcl}
                \mathcal{R}_{sw} & = & |\vv{u}_{sw}|/\omega \\
                \mathcal{R}_{com} & = & |\vv{u}_{com}|/\omega \\
                && \\
                \omega & = & eB\dfrac{n_{sw}m_{sw}+n_{com}m_{com}}{(n_{sw}+n_{com})m_{sw}m_{com}}
                \end{array}
        .\end{equation}\\
        
        Still considering the case where both populations have the same initial parallel velocity, one can also always choose an inertial Cartesian frame in which $\underline{\vv{u}_{sw}}$ is along the $x$-axis, and $\underline{\vv{u}_{com}} = 0$~. This frame is referred to as the comet frame, and is used in Fig. \ref{general}. Both beams describe circles in velocity space with the same angular speed and the same centre $\underline{\vv{v}_i}$. The corresponding motion of the ions in physical space is a trochoid, the most general two-dimensional (2D) gyromotion. In the comet frame, the particles belonging to the cometary population (${com}$) describe a more classical cycloid, as they periodically reach a velocity equal to $\vv{0}$~. In Fig. 2, one can see that as the density ratio increases, $|\vv{v}_i|$ becomes closer to the origin. This can be interpreted as the corresponding slowing down of the plasma fluid for spatial scales much larger than the generalised gyroradius. \\
        
        We note that if neither the ions nor the electrons have a velocity component parallel to the magnetic field, and they will not gain such a component during the interaction. We therefore obtain $\underline{\vv{u}_i} = \underline{\vv{u}_e}$~. More generally, this equality is only verified if at any point in time, electrons and ions have the same parallel velocity, independent of the reference frame.\\
        
        The dynamics depends greatly on the density ratio. If $n_{sw} \gg n_{com}$~, then $\underline{\vv{v}_i} \sim \vv{u}_{sw}$ (top-part in Fig. \ref{general}), and the seldom cometary ions behave as test particles in the almost undisturbed flow of the population $sw$. The cycloid has then a radius equal to the cometary ion Larmor radius. As the density ratio $n_{com}/n_{sw}$ increases, particles of the cometary population still describe a cycloid, though the corresponding radius decreases. When the density ratio is equal to the inverse of the mass ratio, both populations move along cycloids of equal radius, as seen in the middle panel of Fig. \ref{general}. \\
        
         If the beams do not have the same initial parallel velocity, independent of the choice of frame, at least one of the populations will drift along the $y$-axis at a constant speed. This is the case in the comet frame, when accounting for the Parker spiral angle: both populations present an additional drift perpendicular to the $x$-axis, with opposite directions. Flipping the sense of the magnetic field does not change the direction of the drift, and the average Parker spiral configuration of the interplanetary magnetic field induces a dawn-dusk asymmetry in the ion dynamics around the comet. This topic is tackled in \citet{behar2018mnras}, based on statistics over the entire mission, and is based on this generalised gyromotion. \\
        
        This generalised gyromotion is a collisionless 3D description of such a beam-beam interaction, much broader than the following 2D application for a comet.

\section{Generalised gyromotion in a cometary atmosphere}

        We will now see to what extent this generalised gyromotion can describe the dynamics of the solar wind during its interaction with a comet. In order to resolve the equation of motion for the solar wind protons in Eq. (\ref{dynamics}), we need to express three main parameters, namely the cometary ion density and velocity, and the magnetic field.

\subsection{Cometary ion density}

        The spatial distribution of the cometary ions is a major ingredient of the model, as it defines what obstacle is presented to the solar wind. The cometary atmosphere is assumed to have a spherical symmetry. For this exercise, the size of the nucleus is negligible, and so is its mass: the neutral elements, produced at a rate $Q$, are expanding radially in all directions with a constant speed $u_0$~. We assume these particles to be water molecules, $H_2O$. They are ionised or photo-dissociated at a rate $\nu_d$. By writing the equation of continuity with source terms on the cometary neutral density $n_0$~, we obtain, in this spherical symmetry,
        \begin{equation}
                \frac{1}{r^2} \frac{d (r^2 n_0 u_0) }{d r} = -\nu_d \ n_0~,
        \end{equation}
        with the following solution established and used by \citet{haser1957bcs}:
        \begin{equation}
                n_{0} (r) = \frac{ Q }{ 4 \pi u_0 r^2} \cdot e^{\ -r/R_d} \quad \quad ; \quad R_d = u_0/\nu_d~.
        \end{equation}
        
        The cometary ions are created by ionisation of the neutral particles with a rate $\nu_i$~. They have the initial radial velocity $\vv{u}_0$ but will immediately be accelerated by the local electric and magnetic fields, to eventually escape the region of the denser coma. We separate the ionised coma into two different cometary ion components: the new-born cometary ions first, which are the main obstacle to the solar wind, and second the accelerated (or pick-up) cometary ions. The dynamics of the first population is assumed to be trivial: the new-born cometary ions move radially away from the nucleus with the same speed as the neutral molecules. The dynamics of the second, however, is much more complex, driven mostly by the mass-loading mechanism, meaning that the pick-up cometary ions leave the system quicker than they would have ballistically. New born ions become pick-up ions at a rate of $\nu_{ml}$. Accordingly, a destruction term appears in the continuity equation of the new-born cometary ions:

        \begin{equation}
                \frac{1}{r^2} \frac{d (r^2 n_{com} u_{0}) }{d r} = \nu_i \ n_{0} - \nu_{ml} \ n_{com}
        .\end{equation}
        
        With $R_i = u_0/\nu_{i}$ and $R_{ml} = u_0/\nu_{ml}$
        \begin{equation}
                n_{com}(r) = \frac{1}{R_i} \frac{R_d R_{ml}}{R_d-R_{ml}}  \left( 1-e^{-r\ \left(\frac{1}{R_{ml}}-\frac{1}{R_d}\right)} \right) \cdot n_0(r)
                \label{dens}
        .\end{equation}
        
        Three characteristic radii are found in the density profile. In Sect. 3.5, we see that $R_{ml}\ll R_d<R_i$~. For $r<R_{ml}$, that is, before the new born ions are accelerated and neglected, $n_{com}(r<R_{ml}) \propto 1/r$~, a result observed by the {\it Rosetta} mission at comet 67P/CG in the first $\sim 200$ km from the nucleus, and discussed by \citet{edberg2015grl}. At the larger scales that we are interested in,  $ R_{ml} \ll \ell \ll R_d $~, the neutral and the ion densities are proportional, while the exponential term is still negligible. Subsequently, we obtain 
        
        \begin{equation}
                n_{com}(R_{ml} \ll r \ll R_d) = \frac{ \nu_i }{ \nu_{ml} } \frac{ Q }{ 4 \pi \ u_0} \cdot \frac{1}{r^2} \quad \quad [m^{-3}]
                \label{densSimple}
        .\end{equation}

        In this description, we have a steady creation and disappearance of the slow, new-born cometary ions that are constituting the bulk mass of the ionised coma, which interacts electromagnetically with the solar wind.

\subsection{Two-dimensional magnetic pile-up}

        Another important term to model in the equation of motion of the protons is the magnetic field $\vv{B}$ within the coma. We first need to express $\underline{\vv{u}_i}$~, the total velocity of the ion fluid, in which the B-field is considered to remain frozen-in:
        
        \begin{equation}
                \underline{\vv{u}_i} = \underline{\vv{u}_e} = \frac{n_{sw}}{n_{sw} + n_{com}} \vv{u}_{sw} + \frac{n_{com}}{n_{sw}+ n_{com}} \vv{u}_{com}
        .\end{equation}
        
        Our goal is to solve for $\vv{u}_{sw}$~, therefore necessarily some more assumptions have to be made in order to simplify the total ion velocity and remove degrees of freedom in the system.
        
\begin{center}
        \begin{figure}
                \includegraphics[width=\columnwidth]{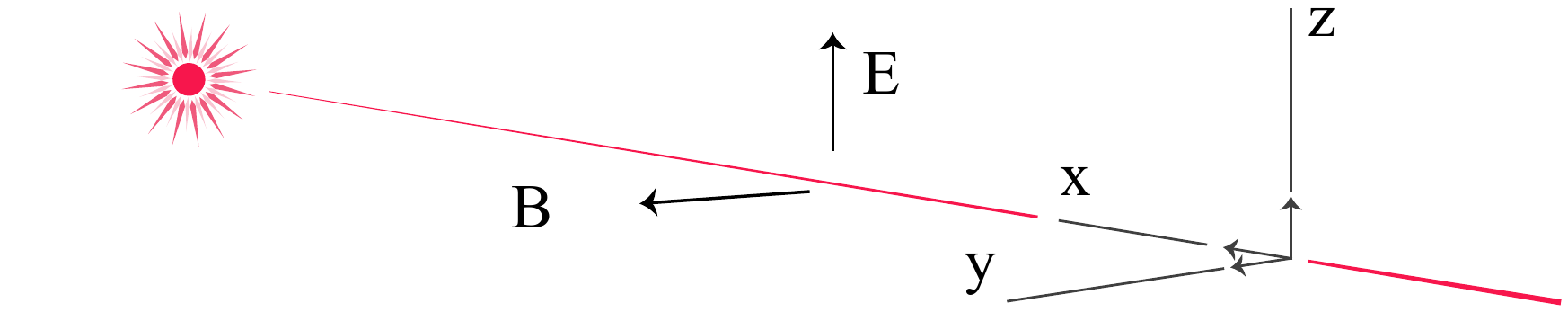}
                \caption{Comet-Sun-Electric field frame of reference. The solar wind dynamics is considered in the plane $y=0$ only. \label{frame}}
        \end{figure}
        \end{center}

        We first define the Comet-Sun-Electric field reference frame, illustrated in Fig. \ref{frame}, as follows: the upstream solar wind magnetic field is directed purely along the $y$-axis, with an amplitude $B_{\infty}$~. The corresponding electric field according to Eq. (\ref{E}) is along the $z$-axis with an amplitude $E_{\infty} = u_{\infty} B_{\infty}$~. In this precise frame (comet-centred CSE), $\vv{u}_{com} \ll \vv{u}_{sw}$~, and it can be shown with the help of Eq. (\ref{densSimple}) that in most of the interaction region, $\underline{\vv{u}_i} \sim \xi_{sw}\vv{u}_{sw}$~. The latter is in agreement with fully kinetic simulations of the interaction~ (e.g. Fig. 2 of~\citealt{deca2017prl}). In the generalised gyromotion, the asymmetry in the total ion velocity appears only because of the different masses of the two populations, as for identical masses $\underline{\vv{u}_i} = \underline{\vv{v}_i}$ (see Sect. 2), that is, both not accelerated and remaining along the $x$-axis only.
                
         As $\vv{B}_{\infty}$ is along the $y$-axis ({\it i.e.} no Parker spiral angle), $\vv{u}_{\infty}$ is along the $x$-axis and as the cometary outflow is spherical, the plane $y=0$ is a plane of symmetry of the system. Therefore, within $y=0$~, neither $\underline{\vv{v}_i}$ nor $\underline{\vv{u}_i}$ can have a component along $ \hat{\vv y}$~(corresponding to the fact that no particle acceleration happens along $\vv{B}$). One more simplification is needed to be able to express the magnetic field. We assume that the total ion velocity remains along the $x$-axis and follows:
        \begin{equation}
                \vv{u}_i = -\xi_{sw} u_{\infty} \ \hat{\vv{x}}
                \label{velEl}
        ,\end{equation}

        with $\xi_{sw}>0$ and $u_{\infty}>0$. From Eq.(\ref{velEl}) of the total ion velocity and Eq.(\ref{E}) of the electric field, one finds that  $ \vv{E} = \left(0, -u_i B_z, u _i B_y \right) $ with $u_i=|\vv{u}_i|>0$. Additionally, the steady state Faraday's law $ \vv{\nabla} \times \vv{E} = \vv{0} $~ states that neither $E_y$ nor $E_z$ vary along $x$. Finally, in the plane of interest ($y=0$):
        \begin{equation}
                \vv{B} = \frac{B_{\infty}}{\xi_{sw}} \ \hat{\vv{y}} \\
        \label{pileUp}
        .\end{equation}
        
        The magnetic field frozen in the ion fluid should -- in the absence of a Hall term -- depend on the motion of both the solar wind and cometary ions. Since the advection of the cometary ions is not solved, the total ion velocity cannot be consistently derived.

\subsection{Solar wind proton dynamics}\label{subs:dyn}

        Considering the new-born cometary ion population, one can assume in the cometary frame that $u_{com} \ll u_{sw}$~. Therefore, using (\ref{dynamics}), (\ref{velEl}) and (\ref{pileUp}), we have
        
        \begin{equation}
                \begin{array}{rcl}
                        \dot{\vv u}_{sw} & = & \dfrac{e \ \xi_{com} B}{m_{sw}} \cdot   \vv{u}_{sw}  \times \hat{\vv y} \\
                        & = &  \dfrac{e \ \nu_i Q B_{\infty}}{4 \pi m_{sw} n_{sw} \nu_{ml}  \ u_0} \cdot \dfrac{1}{r^2}   \cdot   \vv{u}_{sw} \times \hat{\vv y} \ \ [m/s]
                \end{array}
        .\end{equation}
        
        The force experienced by an individual solar wind proton is therefore of the form
        
        \begin{equation}
                \vv{F} = \frac{m_{sw} \ \eta}{r^2} \vv{u}_{sw} \times \hat{\vv y}
        .\end{equation}

        The force is always perpendicular to the proton velocity, with a strength proportional to the inverse of the square distance $1/r^2$~. The equation of motion for protons is then
        
        \begin{equation}
                \boxed{\vv{\ddot{r}} =  \frac{\eta}{r^2} \vv{\dot{r}} \times \hat{\vv y} \quad \quad ; \quad \eta = \frac{e \ \nu_i Q B_{\infty}}{4 \pi \ \nu_{ml} n_{sw} m_{sw} u_0} \quad [m^2/s]}
                \label{eqMotion}
        .\end{equation}
    
    In this description, the solar wind protons do not lose energy and are only gyrating, with a gyroradius function of their distance to the nucleus only. This can also be seen as the motion of charged particles in an effective magnetic field always perpendicular to the plane of the motion, with an amplitude proportional to $1/r^2$. This is the core of the model, the reduced form of the solar wind proton interaction with a coma.\\

   The dynamical system defined by \eqref{eqMotion} for solar wind protons is integrable. Its solutions are thoroughly studied in \citet{saillenfest2018aa};  here, we recall their main features. Let us introduce the polar coordinates $(r,\theta)$ in the $(x,z)$ plane of motion. The dynamical equations rewrite\footnote{The coefficient $k$ used by~\citet{saillenfest2018aa} is equal to $-\eta$.}:
   \begin{equation}
      \left|\hspace{0.1cm}
      \begin{aligned}
         \ddot{r}-r\dot{\theta}^2 &= -\frac{\eta}{ r}\dot{\theta} \\
         r\ddot{\theta}+2\dot{r}\dot{\theta} &= \frac{\eta}{r^2}\dot{r}
      \end{aligned}
      \right.
   ,\end{equation}
   where the dot means derivative with respect to time $t$~. These coupled differential equations imply the conservation of kinetic energy $E$ and a generalised angular momentum $C$ that can be expressed as two characteristic radii:
   \begin{equation}\label{eq:rErC}
      \left|\hspace{0.1cm}
      \begin{aligned}
         r_E &= \frac{|\eta|}{u} \hspace{0.5cm}\text{where}\hspace{0.5cm} u = \sqrt{\dot{r}^2 + r^2\dot{\theta}^2} \\
         r_C &= r\exp(1-r^2\dot{\theta} /\eta)
      \end{aligned}
      \right.
   .\end{equation}
   Their respective values fully determine the trajectory, with a bifurcation occurring at $r_C=r_E$~. As for $1/r^3$ magnetic fields \citep{graef1938jmp}, the solution $(\theta,t)$ can be written as a function of $r$ defined by an integral.
   
   The solar wind particles can be considered as originating from infinity on initially parallel trajectories. Since they all have the same conserved velocity $u=u_{sw}$~, the characteristic radius $r_E$ acts only as a scaling parameter (whereas the particles of the solar wind span all the possible values of $r_C$). With this setting, \cite{saillenfest2018aa} show that around the nucleus at the origin, a circular cavity totally free of particles is created with radius\footnote{The exact radius of the cavity is $r_\mathrm{cav}/r_E=W_0(1/e)$, where $e=\exp(1)$ and $W_0$ is the positive branch of the Lambert $W$ function. Its first decimals are $W_0(1/e)=0.2784645427610738...$} $r_\mathrm{cav}\approx 0.28\,r_E$~.
   
\begin{center}
        \begin{figure*}[h!]
        \includegraphics[width=\textwidth]{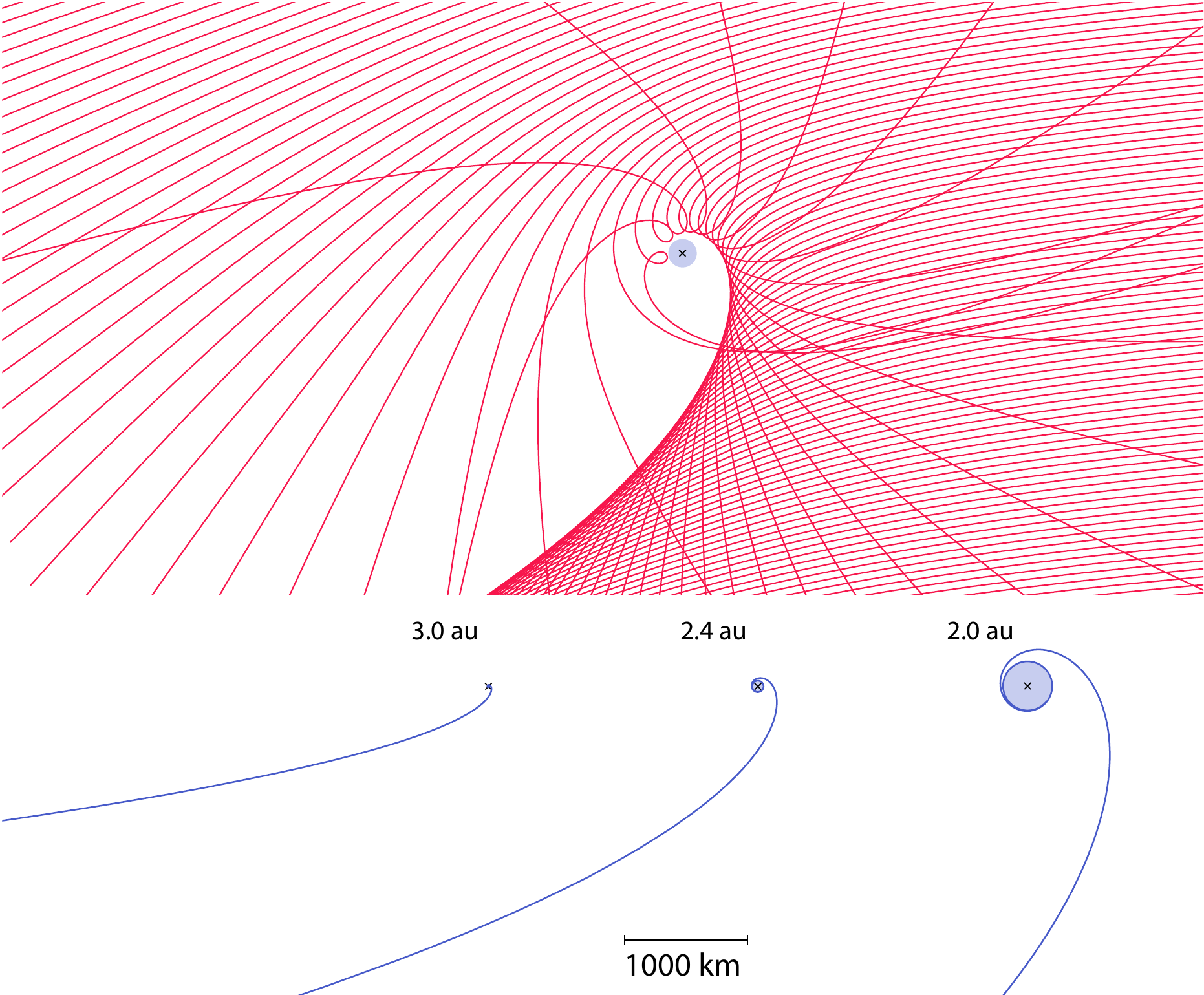}
                \caption{Top row: examples of solar proton trajectories, dimensionless, initially flowing from the right to the left. No particle can enter into the cavity, the central disk of radius $r_\mathrm{cav}\approx 0.28\,r_E$~. Bottom row: the shape of the caustic created by particles coming from infinity, using the same spatial scale for three different heliocentric distances, as developed by \citet{saillenfest2018aa}. The corresponding values of $r_E$ are, from left to right: 27, 165 and 714~km. Near the origin, the caustic wraps around the cavity. The nucleus position is displayed by a black cross in all plots.}
        \label{traj0}
        \end{figure*}
        \end{center}

   The resulting trajectories are shown in Fig. \ref{traj0}. A portion of the incoming flux of particles is temporarily focussed along a very specific curve, defined as the crossing points of infinitely close neighbour trajectories. By analogy to light rays, we call it a ``caustic'', resulting in an overdensity of particles. This caustic has a well-defined shape, which can be expressed as the root of the variational vector. It is plotted in Fig. \ref{traj0} for different values of the scaling parameter $r_E$~, which rescales according to the comet activity and the heliocentric distance. A similar overdensity curve can be observed in other contexts, such as a flux of charged particles in the equatorial plane of a magnetic dipole \citep{Stormer1930,Shaikhislamov2015}, or the deflection of solar particles around the thin atmosphere of Pluto \citep{McComas2008}. This could indicate that analogous processes are at play. Further discussions regarding a general $1/r^n$ law are given by \cite{saillenfest2018aa}.
   
   We note that the notion of ``impact parameter'', $ z_\infty = z(x \rightarrow \infty),$ has no clear meaning for a $1/r^2$ effective magnetic field (it is infinite for every particle). We should therefore express the problem in another way: we simply deal here with a far-enough starting distance for the particles, such that we can safely assume that their trajectories are parallel.

        \subsection{Third dimension}
        
        Outside of the plane ($y=0$), the magnetic field draping introduces an angle between the magnetic field and the normal to the Comet-Sun line. As previously mentioned, such an angle would result in an additional acceleration of the solar wind along the $y$-axis, and the dynamics out of the $(y=0)$-plane would not be planar. Just as its pile-up, the draping of the magnetic field is also given by the bulk velocity of the ions, assuming the field is frozen in the ion flow. Therefore a generalisation of the model to the third spatial dimension would result in the same physics as this 2D approach in the $(y=0)$-plane.

\subsection{Parameters and scales}\label{subs:paramScales}

   \begin{table}
   \caption{Nominal parameters used to get the value of $\eta.$}\label{param}
   \centering
   
   \begin{tabular}{rccc}
   \hline
    & 1 au & 4 au & law  \\
   \hline
    $Q_i$ & $2.6\cdot 10^{28}$ s$^{-1}$& $2.2\cdot 10^{25}$ s$^{-1}$ & $\propto R^{-5.10}$ \\
    $Q_o$ & $1.6\cdot 10^{29}$ s$^{-1}$& $7.8\cdot 10^{24}$ s$^{-1}$ & $\propto R^{-7.15}$ \\
    $\nu_i$ & $6.5 \cdot 10^{-7}$ s$^{-1}$ & $1.8 \cdot 10^{-8}$ s$^{-1}$ & $\propto R^{-2}$ \\
    $\nu_d$ & $1.8 \cdot 10^{-5}$ s$^{-1}$ & $1.1 \cdot 10^{-6}$ s$^{-1}$ & $\propto R^{-2}$ \\
    $n_{\infty}$ & 5.0 cm$^{-3}$ & 0.3 cm$^{-3}$ & $\propto R^{-2}$  \\
    $|\vv{B}|$ & 4.6 nT & 1.2 nT & $\propto (R-\alpha)/R$ \\
    $u_{0}$ & 0.7 km/s & 0.7 km/s & - \\
    $\nu_{ml}$ & 0.01 s$^{-1}$ & 0.01 s$^{-1}$ & -\\
    $\eta$ & $1.7 \cdot 10^{13}$ m$^2$s$^{-1}$ & $3.7 \cdot 10^{9}$ m$^2$s$^{-1}$ & - \\
    \hline
    $u_{\infty}$ & 400 km/s & 400 km/s & - \\
    \hline
    \\
    \\
   \end{tabular}
   \end{table}
   
        \begin{center}
        \begin{figure}
        \includegraphics[width=\columnwidth]{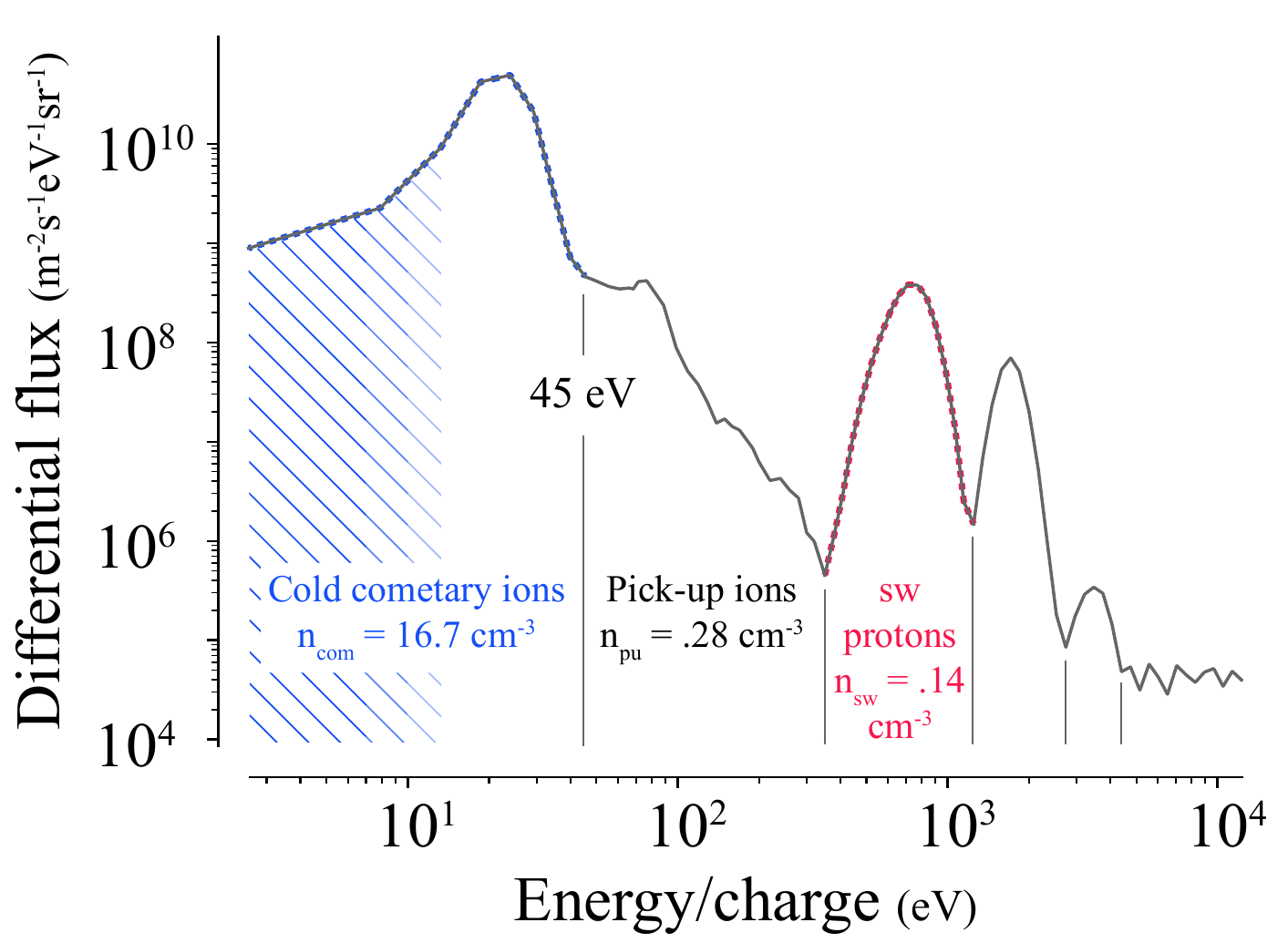}
                \caption{Ion spectrum (differential flux vs. energy per charge of the ions) taken on 2014-11-28 at 2.88 au, integrated for 3 hours. The stripped region for low energies indicates measurements affected by the spacecraft potential.}
        \label{flux}
        \end{figure}
        \end{center}
   
        Table \ref{param} gives the physical quantities in the factor $\eta$ together with their evolution with the heliocentric distance $R$~. At comet 67P/CG, the creation rate of neutral volatiles $Q$ was found to be asymmetric around perihelion, with a higher activity after perihelion. A multi-instrument analysis can be found in \citet{hansen2016mnras}, resulting in an empirical analytical fit given in Table 1. When necessary, we use the notation $Q_i$ for the pre-perihelion in-bound leg of the {\it Rosetta} mission, and $Q_o$ for the post-perihelion out-bound leg. The drawback of this empirical model is a discontinuity in the value of $Q$ at perihelion. The value of the destruction rate $\nu_d$  and the ionisation rate $\nu_i$ and their dependence on the heliocentric distance is taken from \citet{crovisier1989aa}. The magnetic field function of the heliocentric distance can be found in \citet{cravens2004}.\\
        
        The rate $\nu_{ml}$ at which new-born cometary ions are considered to turn into accelerated pick-up ions is arbitrary, but plays an important role in various parameters and scales of the model. It should be much larger than $\nu_{i}$ and $u_0/\ell$~, with $\ell$ being the characteristic length scale of the system. A proxy of its value can be obtained from {\it in situ} data. In a mission overview, using the imaging spectrometer RPC-ICA onboard {\it Rosetta}, \citet{nilsson2017mnras} show that a large majority of cometary ions observed at the spacecraft are rather slow, typically below a few tenths of an  electron Volt. Detaching from this cold population, accelerated cometary ions are observed with fluxes two to three orders of magnitude lower. In the case study of \citet{behar2016grl}, which used data from the RPC-ICA instrument as well, most of the values needed for estimating $\nu_{ml}$ can be found, with the exception of the proton and the cold cometary ion densities. Both densities are given here as the integrated plasma moment of order zero, on the same day. The data were taken on 2014-11-28, a day representative of the general cometary ion spectrum during the mission. An integrated spectrum (differential flux function of the energy per charge of the ions) is given in Fig. \ref{flux}. The probe was 2.88~au from the Sun, 30~km away from the nucleus on a terminator orbit.

        The cold ions reach energies up to 45 eV (22 km.s$^{-1}$), and have an average density of 16.7 cm$^{-3}$. The spacecraft potential, observed  to be negative, on average, and often below $-10$ eV \citep{odelstad2017mnras}, accelerates the positive ions towards the spacecraft, so that the ions collected by the instrument appear more energetic that they actually are in the cometary plasma. The peak in the flux of the cometary new-born ions at about 20 eV corresponds to much lower energies, closer to 0 eV: the population modelled by cometary ions at rest in the analytical model, disappearing with rate $\nu_{ml}$. Unfortunately, no spacecraft potential measurement is available on that day. The solar wind protons have an average speed of 376 km.s$^{-1}$ (734 eV), with an average density of 0.14 cm$^{-3}$. One obtains a total ion fluid velocity of $\xi_{sw}u_{sw} \sim 3.13 $ km.s$^{-1}$. With an average magnetic field of about 15 nT as measured by RPC-MAG that day, the motional electric field is about 0.05 mV.m$^{-1}$. Assuming a constant acceleration, we get a very coarse duration of 77 s for the cometary ions to reach 45 eV~. In summary, after about 77 seconds, slow new-born cometary ions surpass the kinetic energy of 45 eV (22 km.s$^{-1}$) and become `accelerated' cometary ions, or pick-up cometary ions that have a density of 0.28 cm$^{-3}$~. The corresponding rate $\nu_{ml}$ is 0.011 s$^{-1}$ (corresponding at this heliocentric distance to a value of $r_E=77.5$ km). Moreover, on that precise day, it is fair to neglect the accelerated cometary ions, 60 times less dense than the cold ions. For the rest of the study, we consider $\nu_{ml}$ to be on the order of  0.01 s$^{-1}$ in magnitude: new-born cometary ions are neglected after being accelerated for 100 s.\\

\subsection{Particle-field feedback}

        The steady-state assumption together with the scales that are considered in the problem result in extremely simplified Faraday's and Amp\`ere's laws. Close to the caustic, two beams are seen, one incident and one emerging from it. These two beams of identical speed but different direction have the effect of decreasing the local bulk speed. The magnetic field should also be affected, slightly increasing along this structure. In turn, particle trajectories will be corrected by this magnetic feedback. Particles and fields will affect each other until steady state is reached. The model however cannot go further than the third step -- the bulk speed decrease -- in the following sequence.
                
         \includegraphics[width=.9\columnwidth]{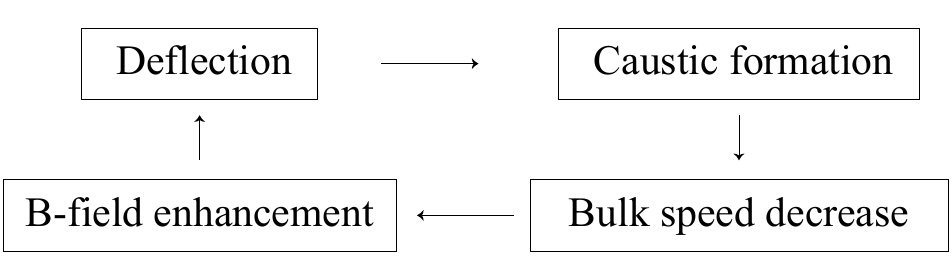}\\
         
         Yet another source of magnetic field pile-up is missing: pick-up cometary ions that gained energy from the interaction are neglected. Even though these ions are present everywhere, they will more significantly increase the total cometary ion density in the +$z$-hemisphere of the coma. The magnetic field will pile-up slightly more in this hemisphere and the proton deflection is expected to be somewhat higher for positive $z$ values, close to the nucleus. An example of such effects can be found in the data and simulation analyses of \citet{koenders2016mnras}.

\subsection{Consistency of the model \& Summary}

        We gather the main assumptions and orderings we have been working with below. We remind that $\ell$ is the characteristic length of this steady-state system, $u$ is its characteristic velocity, and $\ell_P$ is the characteristic length of the electron pressure gradient.
        
        \begin{enumerate}[i]
        \item ~~$\ell \gg d_i$
        \item ~~$\ell_P \gg r_{ge} v_{the}/u$
        \item ~~$\partial_t \cdot \equiv 0$
        \item ~~$\vv{u}_{com} = \underline{\vv{u}_{com}}$ and $\vv{u}_{sw} = \underline{\vv{u}_{sw}}$
        \item ~~$\vv{B} \perp \vv{u}_{sw}$
        \item ~~$R_{ml} \ll \ell \ll R_d < R_i$
        \item ~~$n_{com} \sim n_{new-born}$
        \item ~~$\vv{u}_{com} \ll \vv{u}_{sw}$
        \item ~~$\underline{\vv{u}_i} = \xi_{sw}u_{\infty}\hat{\vv{x}}$
        \end{enumerate}

        Since the phenomenon we model is the deflection of the solar wind, the characteristic length $\ell$ can be considered as the distance over which a particle is deflected by some amount. In other words, the characteristic length is a fraction of the radius of curvature of the trajectories, which in our case has the simple and convenient expression $\rho = -u_{\infty}r^2/\eta$~. We choose the definition $ \ell \equiv \rho$~, corresponding to the distance over which a particle is deflected by 1 radian. The smallest radius being modelled is the radius of the cavity, $r_{cav} \sim 0.28 \ |\eta|/u_{\infty}$~, and therefore the assumptions are not verified closer to the nucleus. 

We now review these assumptions. \\
        
        i ~~~~ $\ell \gg d_i$  ~~ --- ~~ We compared the values of $\ell$ and $d_i= c/e\sqrt{\epsilon_0*m/n} $ depending on the heliocentric distance, at the cavity radius (the most constraining cometocentric distance). For large heliocentric distances and at the cavity, both terms become of the same order of magnitude. Farther away from the nucleus, the ordering is verified. \\
        
        ii ~~~~ $\ell_P \gg r_{ge} v_{the}/u$  ~~ --- ~~  Assuming an isothermal and spherically symmetric coma, the typical length scale of the electronic pressure gradient is $\ell_P = P_e/\partial_rP_e = r(n_{com}+n_{sw})/2n_{com}$. $r_{ge}=m_e v_{the}/(B \ e)$ is the electron gyroradius, $ v_{the} = \sqrt{(2 \ e \ E_e/m_e)}$ is the electron thermal speed, with $E_e\sim7.5 eV$ being the electron thermal energy taken from \citet{eriksson2017aa}. The thermal speed for both the solar wind electrons and the cometary electrons are of the same order of magnitude, which is greater than the average speed of each electron population. Similarly to the Hall term of the electric field, condition ii is not fulfilled at large heliocentric distances and close to the nucleus.  \\
        
        iii ~~~~ $\partial_t \cdot \equiv 0$  ~~ --- ~~  Dynamic phenomenons are not modelled, and the system can only reach another state adiabatically, with changing upstream conditions. \\
        
        iv ~~~~ $\vv{u}_{com} = \underline{\vv{u}_{com}}$ and $\vv{u}_{sw} = \underline{\vv{u}_{sw}}$  ~~ --- ~~  This is one of the strongest assumptions made in the model, that actually allows one to consider protons as single particles: the trajectories can only be relevant if they are not crossing each other. In the flow presented in Fig. \ref{traj0}, as long as protons do not reach the caustic, neighbouring trajectories are never intersecting: only the density and the bulk velocity change, but the assumed beam distribution is not deformed. We note that this is still true on the caustic itself, where, by definition, all trajectories are aligned. However, immediately after the caustic, two beams of comparable density appear in velocity space. This is a very interesting situation that is not accounted for in the present model. After the caustic, the beam quickly looses density and its particles will experience electric and magnetic fields dictated by the `upstream' beam: they will have a general gyromotion which is not modelled here. \\
        
        v ~~~~~ $\vv{B} \perp \vv{u}_{sw}$  ~~ --- ~~  For an upstream magnetic field along the $y$-axis, the symmetry of the system guarantees this configuration everywhere in the plane ($y=0$)~. The average Parker spiral angle will break this symmetry, and the two populations will have an additional drift along the $y$-axis. In a purely parallel case, $\vv{B} = B \ \hat{\vv{x}}$ and no solar wind deflection nor cometary ion acceleration can happen. At heliocentric distances above 1 au~, $\vv{B}$ becomes closer to 90$^\circ$ from the $y$-axis on average, and projected in the plane of interest we expect the dynamics to be qualitatively identical to the one depicted here-above. \\
        
        vi ~~~~ $R_{ml} \ll \ell \ll R_d < R_i$  ~~ --- ~~  In  Sect. 3.5, we have found that $R_{ml} \sim 10^2$ km, $R_d \sim 10^6$ km and $R_{i} \sim 10^8$ km, verifying $R_{ml} \ll R_d < R_i$~. For $r \sim R_d$ and beyond, the cometary ion density is already negligible compared to the solar wind density. We remind that only the ratio of the densities matters in the dynamics, and therefore the change of slope in the density profile at these scales barely has any impact on the solar wind dynamics. \\
        
        vii  ~~~~ $n_{com} \sim n_{new-born}$  ~~ --- ~~  (Discussed with viii )
        
        viii ~~~~ $\vv{u}_{com} \ll \vv{u}_{sw}$  ~~ --- ~~  The observations presented by \citet{nilsson2017mnras} show that over the mission duration and as seen by the spacecraft, these two assumptions are sound. The flux of pick-up cometary ions is always two to three orders of magnitude above the flux of low-energy cometary ions, a difference that cannot be evened out by the difference of speeds (see also Fig. \ref{flux}). The solar wind energy was also never close to the cold cometary ions' energy. \\
        
        ix ~~~~ $\underline{\vv{u}_i} = \xi_{sw}u_{\infty}\hat{\vv{x}}$  ~~ --- ~~ This is actually inconsistent with the generalised gyromotion, as the ion bulk velocity is seen to change its direction (precisely the interest of the model). But as of now, this simplification seems necessary, and allows us to reduce the proton dynamics to the simple object of Eq. (\ref{eqMotion}). We also mentioned that for most cometocentric distances, $\underline{\vv{u}_i} \sim \xi_{sw}\vv{u}_{sw}$~, so as long as the deflection of the protons is not too high, the resulting pile-up is relevantly modelled (within the limits of our description of the coma density). \\
        
        In summary, different aspects of the flow in Fig. \ref{traj0} are to be considered lightly, and these are:
        
        \begin{itemize}
        \item Trajectories of protons some time after they passed the caustic are non-physical (see iv). The local density is however only slightly disturbed.
        \item The magnetic pile-up will be affected in areas where protons have experienced significant deflection (see ix), typically the region downstream of the caustic, that is, the lower-left quadrant of the top graph in Fig. \ref{traj0}. For these two reasons, the solar wind ion cavity should most likely not be circular.
        \item In the region closest to the nucleus at large heliocentric distances, electrons and ions are expected to decouple and pressure gradients will be at work (see i and ii), which is not accounted for by the model.
        \end{itemize}
        
        Finally, the entire subject of waves and instabilities has been voluntarily neglected in the pursuit of simplicity, and it is believable that close to the nucleus and close to the Sun, these phenomena begin to play a major role. We refer to the work of \citet{sauer1996grl}, in which the authors study magneto-acoustic waves propagating transverse to the magnetic field in the frame of the bi-ion fluid theory, in a similar system (the artificial comet experiment conducted by the {\it AMPTE} mission).


\section{{\it Rosetta} data \& self-consistent models in the literature}

        We now have a 2D model to describe the velocity of individual solar wind protons around a comet. The dynamics is governed by a simple force acting on protons that is always perpendicular to their velocity and has an amplitude proportional to $1/r^{2}$~. The resulting flow in Fig. \ref{traj0} is highly asymmetric and is only scaled with varying heliocentric distances. We now briefly review a selection of studies and results that support the model, both from {\it in situ} data and from numerical simulations.
        
        \subsection{{\it Rosetta}}
        
        A force always orthogonal to the solar wind protons and proportional to $1/r^2$ in strength was initially proposed as an empirical model in \citet{behar2017mnras} to account for high solar wind deflection close to the nucleus together with very low deceleration, which eventually lead to the creation of a solar wind ion cavity. The ion cavity and the diamagnetic cavity observed at 67P/CG and reported by \citet{goetz2016aa,goetz2016mnras} are different, the former being much larger than the latter. More detailed and physical differences are discussed by \citet{sauer1994grl} and \citet{behar2017mnras}. It was also pointed out that just before the expanding cavity passed the spacecraft position\footnote{The spacecraft can be considered as standing still, close to the nucleus.}, the deflection was focusing on a value of around 140$^{\circ}$, with proton velocity distributions more stable for a time. This would correspond to the crossing of the caustic, at the vertical ($x=y=0$ and $z>0$ in the CSE frame, Fig. \ref{frame}) of the nucleus.
        
         The orbit of the spacecraft over the two years of active mission provided two opportunities to map the solar wind flow over an extended region. The first was a day-side excursion, which ended up being almost entirely within the solar wind ion cavity and therefore of little interest here. During the second excursion, which was conducted at lower activity and in the night-side of the coma, the spacecraft reached distances up to almost 1000 km, and the solar wind was observed during the entire excursion. In \citet{behar2018aa_ns}, it is shown that a combination of the spacecraft position and of the upstream electric field orientation results in the spacecraft being fairly close to the plane of the model ($y=0$ in the CSE frame) during most of the excursion. Ion data present an excellent match with the modelled flow, especially in the +$z_{CSE}$-hemisphere. The value of $\nu_{ml}$ giving the best fit with the data is 0.01 s$^{-1}$ (with all other parameters taken from other studies), surprisingly close to the value found above, $\nu_{ml}=0.011$ s$^{-1}$. We note that two independent methods based on different ion populations give the same value of $\nu_{ml}$~.
        
        Additionally, no significant deceleration in the night-side of the coma was seen to correlate with the deflection, itself observed from just a few degrees up to 70$^{\circ}$~. As described by Eq. (\ref{eqMotion}) and by the overall mission analysis of \citet{behar2017mnras} and \citet{nilsson2017mnras}, the solar wind mostly gives momentum to the coma, without significant loss of kinetic energy. The present work provides a physical explanation for this important observation, and shows under which assumptions the solar wind can indeed be deflected by any angle, with negligible loss of kinetic energy.

        \subsection{{\it Numerical models}}
        
                \begin{center}
        \begin{figure}[h!]
        \includegraphics[width=\columnwidth]{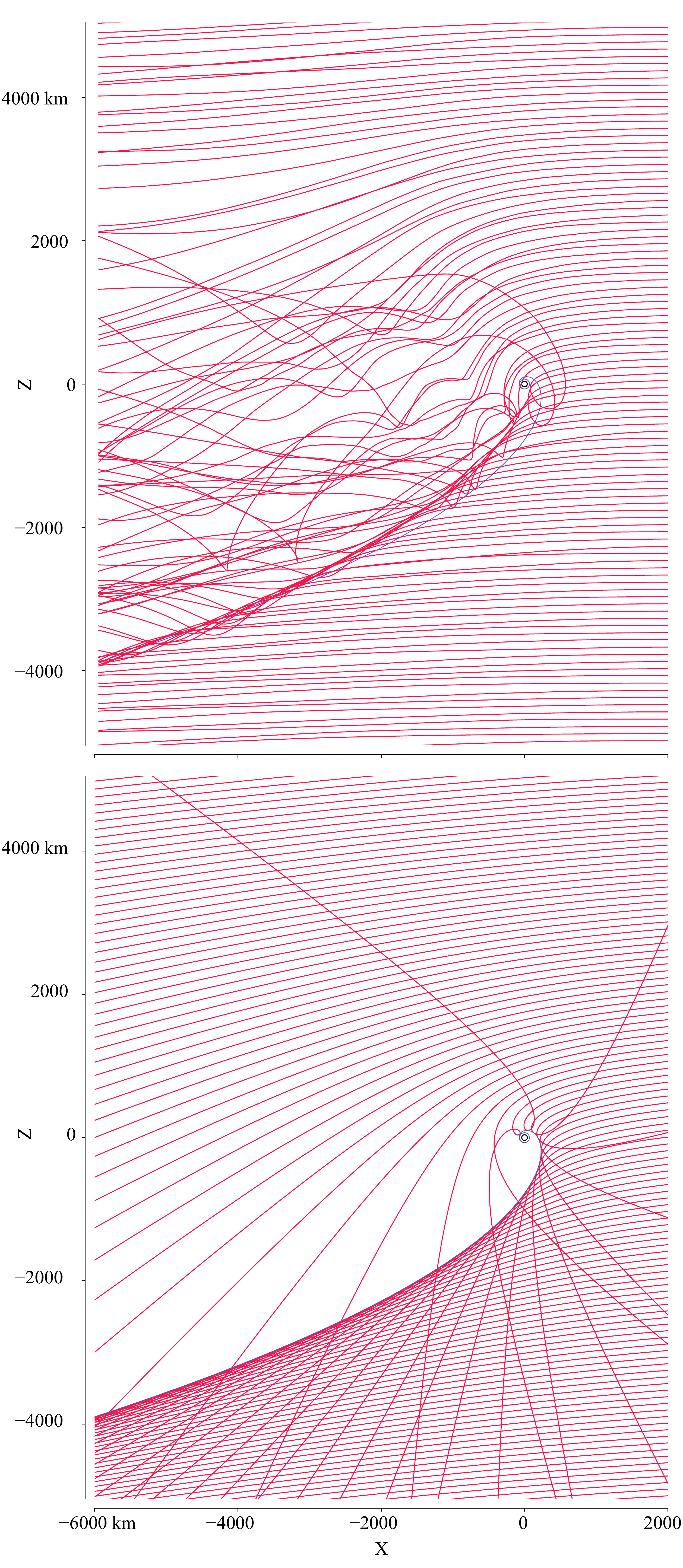}
                \caption{Solar wind proton trajectories (red lines) simulated using hybrid FLASH, a self-consistent numerical model (upper panel) and the trajectories of the 2D semi-analytical model (lower panel). The theoretical position of the caustic is given by the blue line.}
        \label{comparisonFLASH}
        \end{figure}
        \end{center}
        
        The analytical expressions of the generalised gyromotion were verified with the hybrid simulation results, found in \citet{lidstrom2017}.
        
        The solar wind deflection pattern and the corresponding caustic (an over-density structure in the solar wind) can be seen in the results of numerical simulations, in several publications. In the context of comet 67P/CG, this curved over-density in the solar wind can clearly be seen in the results of \citet{wiehler2011asr} (Fig. 3-a), \citet{koenders2016aa} (Fig. 14-a), \citet{koenders2016mnras} (Fig. 3-b), \citet{behar2016aa} (Fig. 6), and \citet{deca2017prl} (Fig. 4-c). Many of these results also show the general deflection of the solar wind, in qualitative agreement with the present model.
        
        Such an asymmetric density structure in the solar particle flow can also be spotted in the simulation of the plasma environment at other solar system bodies. A first example is the solar wind dynamics modelled by \citet{delamere2009jgr} at Pluto; see their Fig. 4. There, for two neutral production rate cases, the flow is highly asymmetric and develops a similar structure along which proton trajectories intersect. An even more familiar result can be found in \citet{kallio2012eps}, with the simulation of the solar wind interaction with unmagnetised bodies like Mars or Venus. Figure 3 presents the effect of mass-loading on the solar wind, in a test run where the the body has no physical extent, and newborn ions are created according to a $1/r^2$ law, with a total production rate of 10$^{26}$ s$^{-1}$~. This is virtually the same system as treated here, and therefore the strong agreement between the flow line of Fig. 3-b in \citet{kallio2012eps} and the proton trajectories modelled here-above is natural.\\
        
        However, all these results present only the bulk velocity of the flow, with the exception of \citet{delamere2009jgr}. This makes it impossible to judge how single particles behave in the structure itself. In Fig. \ref{comparisonFLASH}, trajectories of single solar wind protons taken from a self-consistent numerical model are given as a first illustrative overview. The hybrid-FLASH model is a hybrid particle-in-cell solver\footnote{Ions are treated as massive particles and electrons as a massless charge-neutralising fluid.} developed by \citet{holmstrom2010,holmstrom2013asp}, and used in the context of 67P/CG in \citet{behar2016aa} and \citet{lindkvist2018aa}. The model of the comet was purposefully kept simple, with a spherically symmetrical outgassing (Haser model), two ion populations (solar wind protons and cometary water ions), no charge exchange between neutral particles and ions, and in this particular simulation, $\vv{B}_{\infty}$ is along the $y$-axis. The cell size is constant and equal to 50 km. The heliocentric distance is 2 au, the production rate is $Q = 7.52\cdot 10^{26}$ s$^{-1}$, the ionisation rate is $\nu_i = 1.63\cdot10^{-7}$ s$^{-1}$, and the speed of neutral molecules is 700 m/s. The upstream magnetic field of the  solar wind is along the $y$-axis, $B = 2.53$ nT, its upstream speed is $u_{\infty} = 430$ km/s and its upstream density is $n_{\infty} = 1.25$ cm$^{-3}$.
        
        The value of $\nu_{ml}$ that gives the best match between the caustics, especially at large distances, is found to be $\nu_{ml}=0.025$ s$^{-1}$, which is larger than the previous experimental estimates. We find that similarly to in the semi-analytical model, proton trajectories are slowly deflected, and intersect each other in the (-$x$, -$z$) quadrant, forming a caustic. Immediately after passing the caustic, the protons are now experiencing the electric field mostly dictated by the denser incoming beam, and therefore starts a more complex gyromotion not accounted for by the analytical model. These protons are accelerated upward and cross the caustic. In both the simulation and the analytical model, immediately below the caustic, the phase space distribution function of the solar wind protons presents two beams (which might be similar to the observations of two proton beams reported by \citet{jones1997asr} at comet Grigg-Skjellerup). One can also note that the presence of the pick-up ions, denser in the +$z$ hemisphere, can be seen in the locally higher deflection of the solar wind protons. We do not discuss the situation for cometocentric distances below 500 km, as with too few cells, one cannot properly resolve the smaller scales in this inner region, where charge exchange is also expected to play a role. A noteworthy observation is that in the simulation, the finite size of the box leads to an underestimation of the deflection, as can be seen at the boundary 2000 km upstream of the nucleus. This is due to an injection of solar wind protons at the upstream boundary with an initial velocity along the $x$-axis, whereas in the analytical model, protons have already experienced a significant deflection at this cometocentric distance. This issue is pointed out and quantified in \citet{saillenfest2018aa} (\textit{cf.} Sect. 3.2). It is also probably one of the reasons why the best-fit value of $\nu_{ml}$ is larger here, since it must compensate for the distortion induced by the finite size of the simulation box.
        
        Downstream of the caustic, the proton density drops in both panels, and accordingly to Eqs. (\ref{E}) and (\ref{velIon}), so does the electric field. In turn, the newborn cometary ions will be less accelerated than immediately upstream from the caustic, and will accumulate: the caustic shields the newborn cometary ions, and in turn a discontinuity in their density is expected to form along the caustic. This issue goes beyond the scope of this article, and is left for further studies.
        
        Finally, the nature of this structure may now be discussed under a new light. \citet{bagdonat2002emp} describe the structure (the caustic in our description) as one side of an asymmetric Mach cone formed by the front wave of propagating density and magnetic field disturbances, induced by the obstacle -- the newborn cometary ions in this case -- in the incident flow. The complete asymmetry of the cone is however not further discussed. It is extremely interesting to note that the present model does not consider the super-magnetosonic character of the solar wind, nor does it propagate any type of disturbance. Based on this 2D approach, this asymmetric structure is not formed by a propagating perturbation, and therefore is not a Mach cone. Furthermore, the cometary newborn ions are only indirectly the obstacle in this picture, as the over-density (the caustic) is formed purely by the geometry of the deflected flow. The solar wind is forming an obstacle to itself, an obstacle with a shape independent of the magnetosonic Mach number. Developing the model to a third dimension, comparing it more thoroughly to self-consistent numerical models, and studying the effect of the plasma pressure on this structure is however necessary to conclude on this topic.

\section{Concluding remarks}

        We have shown how momentum and energy are transferred between two collisionless plasma beams for spatial scales that are large compared to the ion inertial length and in the case of negligible electron pressure gradients. This so-called generalised gyromotion applies to the most general, or arbitrary, 3D configuration of two plasma beams. \\
        
        There are two possible ways to consider the exchange of energy between the solar wind and the coma. The first way states that at scales that are large compared to the gyroradius, and based on classical fluid concepts, there is necessarily a loss of kinetic energy in the solar wind. The second way considers individual particles of the solar wind, and the present model shows how and under which conditions and assumptions these particles do not lose kinetic energy, at zero order. However, part of the plasma is accelerated in this interaction (pick-up ions); therefore at a higher order, also the solar wind loses kinetic energy. We further note that translated into bulk properties, this model also displays a deceleration of the fluid.\\
        
         In the plane of symmetry of the classical magnetic field draping at comets, the exchange of momentum through the fields between the solar wind and a comet atmosphere results in a very simple expression of the force applied to the protons: this force is perpendicular to their velocity, with an amplitude proportional to $1/r^2$~. The solar particle flow is reduced to a peculiar and highly asymmetric pattern, exhibiting a caustic, which is also seen in numerical models. {\it In situ} data from the {\it Rosetta} mission show strong support to the semi-analytical model, in terms of deflection and speed, and with the observation of a solar wind ion cavity. We note that these results can be straightforwardly extended to solar wind alpha particles.
         
          The cheapness of the model will allow for an extended and systematic comparison with \textit{in situ} data, allowing us to distinguish how dominant the motional electric field is during a given activity level. The model may also point to previously unnoticed structures, such as the peculiar distribution function of the solar wind proton close to the caustic. Furthermore, the model should also greatly ease the understanding of complex mass-loaded solar wind kinetic simulations. \\
   
   Finally, the validity of the model close to the nucleus is expected to crumble closer to the Sun, as a bow shock might form, at least for strong-enough cometary outgassing. There, waves and instabilities will provide additional ways to transfer energy and momentum, and these phenomena will most likely act to transform the described caustic into a bow shock, beyond the domain of validity of this model. Pinning down the conditions for which this transition happens, together with the micro-physics involved, is an obvious direction to explore. However, further out than the potential bow shock, at comets and at unmagnetised bodies as well, the mass-loading is affecting the solar wind flow, and the present model remains relevant.

\begin{acknowledgements}

This work was supported  by the Swedish National Space Board (SNSB) through grants 108/12, 112/13, 96/15 and 201/15. The work at LPC2E/CNRS was supported by ESEP, CNES and by ANR under the financial agreement ANR-15-CE31-0009-01. This work was supported in part by NASA's Solar System Exploration Research Virtual Institute (SSERVI): Institute for Modelling Plasmas, Atmosphere, and Cosmic Dust (IMPACT). Partial support is also acknowledged by Contract No. JPL-1502225 at the University of Colorado from Rosetta, which is an European Space Agency (ESA) mission with contributions from its member states and NASA. The hybrid solver is part of the openly available FLASH code and can be downloaded from http://flash.uchicago.edu/ developed by the DOE NNSA-ASC OASCR Flash Center at the University of Chicago. The hybrid simulations were conducted using resources provided by the Swedish National Infrastructure for Computing (SNIC) at the High Performance Computing Center North (HPC2N), Ume\aa University, Sweden. Part of this work was inspired by discussions within International Teams 336: "Plasma Surface Interactions with Airless Bodies in Space and the Laboratory" and 402: "Plasma Environment of Comet 67P after Rosetta" at the International Space Science Institute, Bern, Switzerland.

\end{acknowledgements}

%
%

\bibliography{cometLib}
\bibliographystyle{aa}

\end{document}